\documentclass[transmag,letterpaper]{IEEEtran}

\usepackage{cite}
\usepackage{amsmath,amssymb,amsfonts}
\usepackage{graphicx}
\usepackage{textcomp}
\usepackage{xcolor}
\usepackage{xspace}
\usepackage{comment}
\usepackage{graphicx}
\usepackage{textcomp}
\usepackage{amsmath}
\usepackage{amssymb}
\usepackage{xcolor}
\usepackage{booktabs}
\usepackage{algorithm}
\usepackage{algpseudocode}
\usepackage{float}
\usepackage{enumitem}
\usepackage{hyperref}
\usepackage[left=0.635in,right=0.635in,top=0.76in,bottom=1.05in]{geometry}
\usepackage[caption=false,font=footnotesize]{subfi
g}
\usepackage[nohyperlinks, printonlyused, nolist]{acronym}
\begin{acronym}
\acro{imsi}[IMSI]{International Mobile Subscriber Identity}
\acro{tmsi}[TMSI]{Temporary Mobile Subscriber Identity}
\acro{supi}[SUPI]{Subscription Permanent Identifier}
\acro{suci}[SUCI]{Subscription Concealed Identifier}
\acro{guti}[GUTI]{Global Unique Temporary Identifier}
\acro{lte}[LTE]{Long Term Evolution}
\acro{ran}[RAN]{Radio Access Network}
\acro{3gpp}[3GPP]{3rd Generation Partnership Project}
\acro{mac}[MAC]{Message Authentication Code}
\acro{gsma}[GSMA]{Global System for Mobile Communications Association}
\acro{sqn}[SYN]{Sequence Number}
\acro{rrc}[RRC]{Radio Resource Control}
\acro{aka}[AKA]{Authentication and Key Agreement}
\acro{autn}[AUTN]{Authentication Token}
\acro{auts}[AUTS]{Synchronization Failure Parameter}
\acro{geran}[GERAN]{GSM EDGE Radio Access Network}
\acro{utran}[UTRAN]{Universal Terrestrial Radio Access Network}
\acro{nr}[NR]{New Radio}
\acro{e-utran}[E-UTRAN]{Evolved Universal Terrestrial Radio Access Network}
\acro{enb}[eNB]{Evolved Node B}
\acro{gnb}[gNB]{GNodeB}
\acro{bs}[BS]{Base Station}
\acro{ue}[UE]{User Equipment}
\acro{imei}[IMEI]{International Mobile Equipment Identifier}
\acro{gsm}[GSM]{Global System for Mobile Communications}
\acro{umts}[UMTS]{Universal Mobile Telecommunications System}
\acro{eps}[EPS]{Evolved Packet System}
\acro{epc}[EPC]{Evolved Packet Core}
\acro{mme}[MME]{Mobility Management Entity}
\acro{as}[AS]{Access Stratum}
\acro{nas}[NAS]{Non-Access Stratum}
\acro{nsa}[NSA]{Non-Stand Alone}
\acro{ims}[IMS]{IP Multimedia Subsystem}
\acro{auc}[AuC]{Authentication Center}
\acro{hlr}[HLR]{Home Location Registers}
\acro{lai}[LAI]{Location Area Identity}
\acro{pki}[PKI]{Public Key Infrastructure}
\acro{iot}[IoT]{Internet of Things}
\acro{cs}[CS]{Circuit Switched}
\acro{ps}[PS]{Packet Switched}
\acro{sim}[SIM]{Subscriber Identity Module}
\acro{sdr}[SDR]{Software-defined Radio}
\acro{uicc}[UICC]{Universal Integrated Circuit Card}
\acro{cots}[COTS]{Commercial-Off-The-Shelf}
\acro{c-rnti}[C-RNTI]{Cell Radio Network Temporary Identity}
\acro{lfm}[LFM]{Link Failure Message}
\acro{ue}[UE]{User Equipment}
\acro{zitis}[ZITiS]{Zentrale Stelle für Informationstechnik im Sicherheitsbereich}
\acro{imsi}[IMSI]{International Mobile Subscriber Identity}
\acro{tmsi}[TMSI]{Temporary Mobile Subscriber Identity}
\acro{lte}[LTE]{Long Term Evolution}
\acro{epc}[EPC]{Evolved Packet Core}
\acro{enb}[eNB]{\ac{eutran} Node B}
\acro{eps}[EPS]{Evolved Packet System}
\acro{mme}[MME]{Mobility Management Entity}
\acro{pcrf}[PCRF]{Policy and Charging Rules Function}
\acro{mcc}[MCC]{Mobile Country Code}
\acro{mnc}[MNC]{Mobile Network Code}
\acro{sgw}[SGW]{Serving-Gateway}
\acro{pdngw}[PDN-GW]{Packet Data Network Gateway}
\acro{hss}[HSS]{Home Subscriber Service}
\acro{cli}[CLI]{Command Line Interface}
\acro{sim}[SIM-Karte]{Subscriber Identity Module}
\acro{pin}[PIN]{Personal Identification Number}
\acro{puk}[PUK]{PIN Unblocking Key}
\acro{nas}[NAS]{Non Access Stratum}
\acro{tkue}[TKÜ]{Telekommunikationsüberwachung}
\acro{eutran}[E-UTRAN]{Evolved UMTS Terrestrial Radio Access Network}
\acro{enodeb}[eNode-B]{Evolved Node B}
\acro{sdr}[SDR]{Software defined radio}
\acro{json}[Json]{JavaScript Object Notation}
\acro{earfcn}[EARFCN]{E-UTRA Absolute Radio Frequency Channel Number}
\acro{ide}[IDE]{Integrated Development Environment}
\acro{soc}[SoC]{System On Chip}
\acro{3gpp}[3GPP]{Third Generation Partnership Project}
\acro{gui}[GUI]{Graphical User Interface}
\acro{ml}[ML]{Machine-Learning}
\acro{ucs2}[UCS2]{Universal Character Set 2}
\acro{isr}{Idle Mode Signaling Reduction}
\acro{os}[OS]{Operating System}
\acro{imsi}[IMSI]{International Mobile Subscriber Identity}
\acro{tmsi}[TMSI]{Temporary Mobile Subscriber Identity}
\acro{supi}[SUPI]{Subscription Permanent Identifier}
\acro{suci}[SUCI]{Subscription Concealed Identifier}
\acro{guti}[GUTI]{Global Unique Temporary Identifier}
\acro{lte}[LTE]{Long Term Evolution}
\acro{ran}[RAN]{Radio Access Network}
\acro{3gpp}[3GPP]{3rd Generation Partnership Project}
\acro{mac}[MAC]{Message Authentication Code}
\acro{gsma}[GSMA]{Global System for Mobile Communications Association}
\acro{sqn}[SYN]{Sequence Number}
\acro{rrc}[RRC]{Radio Resource Control}
\acro{aka}[AKA]{Authentication and Key Agreement}
\acro{autn}[AUTN]{Authentication Token}
\acro{auts}[AUTS]{Synchronization Failure Parameter}
\acro{geran}[GERAN]{GSM EDGE Radio Access Network}
\acro{utran}[UTRAN]{Universal Terrestrial Radio Access Network}
\acro{nr}[NR]{New Radio}
\acro{e-utran}[E-UTRAN]{Evolved Universal Terrestrial Radio Access Network}
\acro{enb}[eNB]{Evolved Node B}
\acro{gnb}[gNB]{GNodeB}
\acro{bs}[BS]{Base Station}
\acro{ue}[UE]{User Equipment}
\acro{imei}[IMEI]{International Mobile Equipment Identifier}
\acro{gsm}[GSM]{Global System for Mobile Communications}
\acro{umts}[UMTS]{Universal Mobile Telecommunications System}
\acro{eps}[EPS]{Evolved Packet System}
\acro{epc}[EPC]{Evolved Packet Core}
\acro{mme}[MME]{Mobility Management Entity}
\acro{as}[AS]{Access Stratum}
\acro{nas}[NAS]{Non-Access Stratum}
\acro{nsa}[NSA]{Non-Stand Alone}
\acro{ims}[IMS]{IP Multimedia Subsystem}
\acro{auc}[AuC]{Authentication Center}
\acro{hlr}[HLR]{Home Location Registers}
\acro{lai}[LAI]{Location Area Identity}
\acro{pki}[PKI]{Public Key Infrastructure}
\acro{iot}[IoT]{Internet of Things}
\acro{cs}[CS]{Circuit Switched}
\acro{ps}[PS]{Packet Switched}
\acro{sim}[SIM]{Subscriber Identity Module}
\acro{sdr}[SDR]{Software-defined Radio}
\acro{uicc}[UICC]{Universal Integrated Circuit Card}
\acro{rtt}[RTT]{Round-Trip-Time}
\acro{cots}[COTS]{Commercial-Off-The-Shelf}
\acro{c-rnti}[C-RNTI]{Cell Radio Network Temporary Identity}
\acro{lmf}[LFM]{Location Management Function}
\acro{sba}[SBA]{Service-Based Architecture}
\acro{sbi}[SBI]{Service-Based Interface}
\acro{amf}[AMF]{Access and Mobility Function}
\acro{upf}[UPF]{User-Plane Function}
\acro{nef}[NEF]{Network Expose Function}
\acro{sepp}[SEPP]{Secure Edge Protection Proxy}
\acro{dn}[DN]{Data Network}
\acro{nrf}[NRF]{Network Repository Function}
\acro{smf}[SMF]{Session Management Function}
\acro{pfcp}[PFCP]{Packet Forward Control Protocol}
\acro{plmn}[PLMN]{Public Land Mobile Network}
\acro{udm}[UDM]{Unified Data Management}
\acro{qos}[QoS]{Quality of Service}
\acro{pws}[PWS]{Public Warning System}
\acro{sib}[SIB]{System Information Broadcast}
\acro{pei}[PEI]{Permanent Equipment Identifier}
\acro{smc}[SMC]{Security Mode Command}
\acro{tls}[TLS]{Transport Layer Security}
\acro{ip}[IP]{Internet Protocol}
\acro{ftp}[FTP]{File Transfer Protocol}
\acro{pdcp}[PDCP]{Packet Data Convergence Protocol}
\acro{gtp-c}[GTP-C]{Generic Tunnel Protocol-Control Plane}
\acro{http}[HTTP]{Hyper Text Transfer Protocol}
\acro{udp}[UDP]{User Datagram Protocol}
\acro{pcf}[PCF]{Policy Control Function}
\acro{ausf}[AUSF]{Authentication Server Function}
\acro{af}[AF]{Application Function}
\acro{nf}[NF]{Network Function}
\acro{uicc}[UICC]{Universal Integrated Circuit Card}
\acro{aes}[AES]{Advanced Encryption Standard}
\acro{rsa}[RSA]{Rivest–Shamir–Adleman}
\acro{seaf}[SEAF]{Security Anchor Function}
\acro{li}[LI]{Lawful Interception}
\acro{csp}[CSP]{Communication Service Provider}
\acro{csps}[CSPs]{Communication Service Providers}
\acro{lea}[LEA]{Law Enforcement Agency}
\acro{leas}[LEAs]{Law Enforcement Agencies}
\acro{poc}[PoC]{proof-of-concept}
\acro{he}[HE]{Homomorphic Encryption}
\acro{ir}[IR]{Information Retrieval}
\acro{pir}[PIR]{Private Information Retrieval}
\acro{wpir}[WPIR]{Weakly-Private Information Retrieval}
\end{acronym}
\algrenewcommand\algorithmicrequire{\textbf{Input:}}
\algrenewcommand\algorithmicensure{\textbf{Output:}}

\def \sys {\texttt{P3LI5}}
\def \proto {\texttt{SparseWPIR}}
\def \he {$\mathcal{E}$}
\def \pir {$\mathcal{P}$}

\makeatletter 
\newcommand{\linebreakand}{%
  \end{@IEEEauthorhalign}
  \hfill\mbox{}\par
  \mbox{}\hfill\begin{@IEEEauthorhalign}
}
\makeatother 

\begin{document}
\IEEEpubid{0000--0000/00\$00.00˜\copyright˜2023 IEEE
}
\title{P3LI5: Practical and confidEntial Lawful Interception on the 5G core}

\author{\IEEEauthorblockN{1\textsuperscript{st} Francesco Intoci}
\IEEEauthorblockA{\textit{EPFL}\\\textit{CYD - Armasuisse} \\
Lausanne, Switzerland \\
francesco.intoci@alumni.epfl.ch\\}
\and
\IEEEauthorblockN{2\textsuperscript{nd} Julian Sturm\\3\textsuperscript{rd} Daniel Fraunholz}
\IEEEauthorblockA{\textit{ZITiS}\\
Munich, Germany \\
name.surname@zitis.bund.de\\}
\and
\IEEEauthorblockN{4\textsuperscript{th} Apostolos Pyrgelis}
\IEEEauthorblockA{\textit{EPFL} \\
Lausanne, Switzerland\\
apostolos.pyrgelis@epfl.ch\\}
\and
\IEEEauthorblockN{5\textsuperscript{th} Colin Barschel}
\IEEEauthorblockA{\textit{CYD} - \textit{Armasuisse}\\
Lausanne, Switzerland \\
colin.barschel@armasuisse.ch}
}

\maketitle

\begin{abstract}
\ac{li} is a legal obligation of \ac{csps} to provide interception capabilities to \ac{leas} in order to gain insightful data from network communications for criminal proceedings, e.g., network identifiers for tracking suspects. With the privacy-enhancements of network identifiers in the 5th generation of mobile networks (5G), \ac{leas} need to interact with \ac{csps} for network identifier resolution. This raises new privacy issues, as untrusted \ac{csps} are able to infer sensitive information about ongoing investigations, e.g., the identities of their subscribers under suspicion. In this work, we propose \sys{}, a novel system that enables \ac{leas} to privately query \ac{csps} for network identifier resolution leveraging on an information retrieval protocol, \proto{}, that is based on private information retrieval and its weakly private version. As such, \sys{} can be adapted to various operational scenarios with different confidentiality or latency requirements, by selectively allowing a bounded information leakage for improved performance. We implement \sys{} on the 5G LI infrastructure using well known open-source projects and demonstrate its scalability to large databases while retaining low latency. To the best of our knowledge, \sys{} is the first proposal for addressing the privacy issues raised by the mandatory requirement for \ac{li} on the 5G core network.

\end{abstract}

\IEEEpubidadjcol
\section{Introduction}
\label{sec:intro}


Lawful interception (\ac{li}) is a statutory requirement for communication service providers (\ac{csps}) to provide authorized law enforcement agencies (\ac{leas}) with the capability to intercept network communications and gain insightful information regarding criminal and anti-terrorism investigations from network data. While in previous generations of mobile networks, \ac{leas} relied solely on radio monitoring techniques, e.g., IMSI-catchers~\cite{PALAMA2021108137}, to intercept network identifiers, this is no longer possible in the new 5G core network~\cite{3gpp.21.915}, where the subscriber long-term identifiers are transmitted in concealed form to protect users' privacy. As a result, the 3rd Generation Partnership Project defines a set of interfaces and standards which allow \ac{leas} to resolve short-term identifiers with the collaboration of the \ac{csps}~\cite{3gpp.33.126, 3gpp.33.127, 3gpp.33.128}, which hold the associations between the permanent and short-term network identifiers.

However, concealing network identifiers while at the same time requiring \ac{csps} to provide LI capabilities contradicts the enhancement of subscriber privacy in 5G, as an untrusted CSP can infer sensitive information by observing the LI interface: On the one hand, the confidentiality of the LEA operations is at stake, as the CSP can obtain more information about the context of current investigations. On the other hand, there is a breach of users' privacy, as the CSP can gain additional information on its subscribers with respect to their involvement in criminal investigations. This raises the need for designing mechanisms that protect the confidentiality of the LI interface on the 5G core.

\IEEEpubidadjcol
We introduce \sys{}, a novel system for confidential \ac{li} on the 5G core. \sys{} employs \ac{pir} (and its relaxed and more performant variant \ac{wpir}), a cryptographic protocol which enables private data retrieval from untrusted databases. Central to \sys{} is \proto{}, a novel information retrieval scheme empowering clients (\ac{leas}) to control information leakage by their queries towards \ac{csps}, thus accommodating diverse privacy and performance needs. Our contributions encompass: (i) \proto{}: A dynamic information retrieval scheme enabling clients to manage information sharing in queries, effectively balancing privacy and performance; (ii) A rigorous privacy analysis employing established metrics in the literature to quantify information leakage; (iii) The development and \ac{poc} of \sys{}, the first system for practical and confidential \ac{li} on the 5G core.

The rest of the paper is organized as follows: We first introduce the necessary background knowledge on the 5G system and the cryptographic primitives upon which we base our solution (\ref{sec:background}). Then, we discuss related work (\ref{sec:related}), and we formally present our system and threat models (\ref{sec:sys_threat_model}). We introduce \proto{}, our generic \ac{ir} scheme (\ref{sec:sparseWPIR}) and we show how to integrate it in the 5G \ac{li} architecture by building \sys{} (\ref{sec:design}). We experimentally evaluate \sys{} (\ref{sec:evaluation}) and we conclude (\ref{sec:conclusion}).

\section{Background}\label{sec:background}

We highlight the main concepts regarding the 5G core~\cite{3gpp.21.915} (\ref{sub:5gs}) and its LI architecture (\ref{sub:li_5g}). Then, we provide background for the cryptographic primitives employed by \sys{} (\ref{sub:crypto}).

\subsection{The 5G System}\label{sub:5gs}

\subsubsection{Registration Procedure}\label{subsub:5g_identifiers}

In normal mode of operations, upon the first registration to the network, \ac{ue} identify to the network via a \textit{Registration Request} with a concealed identifier called \ac{suci}. 
This is an encryption of the \ac{supi}, a unique identifier per SIM card generated by the \ac{csp}. Encryption employs \textit{ECIES} in such a way that only the CSP can recover the SUPI~\cite{3gpp.33.501}. Moreover, the SUCI is formed with fresh randomness, thwarting tracking attacks. Post initial registration, each \ac{ue} receives a fresh random identifier, 5G-\ac{guti} (or in shortened form 5G-\ac{tmsi}), again unfit for \ac{li} due to its randomness.

\subsubsection{5G Core}\label{subsub:5g_core}

The 5G core is the back-bone network of the 5G architecture defined in the 3GPP specification \textit{TR 21.915}~\cite{3gpp.21.915}. It is composed by a plethora of entities called \textit{Network Functions} (NFs), i.e., components offering specific services to the network. In the context of LI, a relevant NF is the \ac{amf}~\cite{3gpp.29.518}, which handles the registration and mobility of the subscribers in the network (e.g., it allocates identifiers after every registration or when a subscriber changes base station).

\subsection{5G Lawful Interception Architecture}\label{sub:li_5g}

\begin{figure}[t]
\vspace{0.1in}
\centering
\includegraphics[width=0.59\columnwidth]{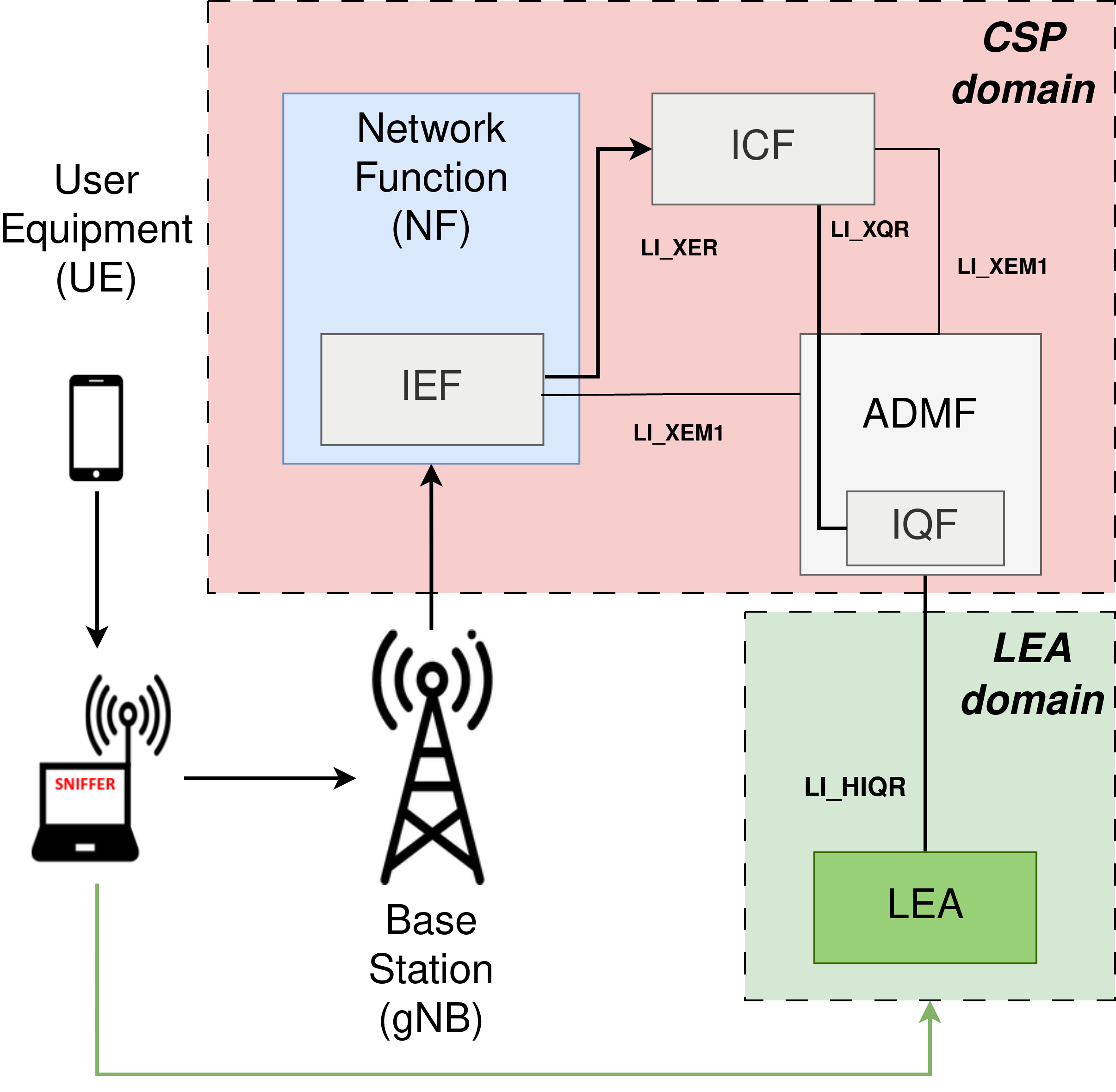}
\caption{LI architecture on the 5G core network for identity disclosure.}
\label{fig:backg_li_arch}
\end{figure}

3GPP specifications~\cite{3gpp.33.126, 3gpp.33.127, 3gpp.33.128, 3gpp.03.120, 3gpp.03.221} define a set of components that CSPs should deploy to provide LI capabilities to LEAs, and the requirements these should satisfy. According to~\cite{3gpp.33.126}, the CSP LI infrastructure should provision the interception of communication upon receiving a lawful request by LEAs. In this work, we consider LI operations aimed at identity disclosure (i.e., by associating temporary identifiers to the SUPI). We briefly describe the main components for target identification in the LI infrastructure (Figure~\ref{fig:backg_li_arch}).

\subsubsection{LEA}\label{subsub:lea}

LEAs deploy sniffing devices to collect identifiers from the network interaction between \ac{ue} and the 5G core targeting the Registration Procedure (\ref{subsub:5g_identifiers}). LEAs communicate with the CSP using a standardized interface named LI\_HIQR. This interface allows the exchange of information from LEA to CSP (e.g., captured identifiers, time and location of the capture, and optionally warrants) and vice versa (status of the interception, resolved identifiers, validity start/end time for the associated identifiers) in the form of XML messages following~\cite{3gpp.03.120, 3gpp.33.128}.

\subsubsection{CSP}\label{subsub:csp}

The LI infrastructure deployed on the 5G Core network by the service provider consists of several components and interfaces which enable communications between them:

\paragraph{IEF}\label{par:ief}
The \textit{Identifier Event Function} is responsible for providing association and disassociation events between SUPIs and short-term (or concealed) identifiers, including the association validity start and end time, to the ICF (\ref{par:icf}) via the LI\_XER interface. The IEF is typically deployed as a separate component (e.g., a virtualized container) associated to a NF (e.g., AMF). 

\paragraph{ICF}\label{par:icf}
The \textit{Identifier Caching Function} is responsible for caching all the identifier associations provided by the IEF. The ICF responds to identity association queries (short-term to long-term identifiers and vice versa) forwarded by the IQF (\ref{par:iqf}). It retains currently valid associations until these are flagged by the IEF as invalid (retaining them in memory for a short time period), or for a maximum retention period (defined by local legislation and the CSP)~\cite{3gpp.33.127}.

\paragraph{IQF}\label{par:iqf}
The \textit{Identifier Query Function} listens for LEA requests on the LI\_HIQR interface, and forwards identity disclosure queries to the ICF via the LI\_XQR interface. It is part of the \textit{Administration Function} (ADMF), which manages the other LI components (i.e., the IEF and ICF) via an internal interface (LI\_XEM1).

\subsection{Cryptographic Building Blocks}\label{sub:crypto}

\subsubsection{Homomorphic Encryption (HE)}\label{subsub:backg_he}

HE schemes support the execution of operations (e.g., addition or multiplications) directly on encrypted data, hence enabling computation delegation to third parties without risking data confidentiality. More formally, let \he{} be an encryption scheme with the following operations: \texttt{KeyGen} which generates an encryption key $sk$, $\texttt{Enc}_{sk}(pt)$ which encrypts a plaintext $pt$ with $sk$, and $\texttt{Dec}_{sk}(ct)$ which decrypts a ciphertext $ct$ with $sk$. \he{} is considered an HE scheme, if there exists an efficient and secure (with respect to a security parameter $\lambda$) algorithm \texttt{Eval}, which takes a circuit $C$ (i.e., a set of operations) and some ciphertexts $[ct_1,\dots,ct_n] {=} [\text{\texttt{Enc}}_{sk}(pt_1),\dots,\text{\texttt{Enc}}_{sk}(pt_n)]$, such that:
\begin{align}
\texttt{Dec}_{sk}(\texttt{Eval}(C,[ct_1,\dots,ct_n])) = C([pt_1,\dots,pt_n])\label{eq:he_def}
\end{align}

Modern HE schemes~\cite{gentry09,FV12,Brakerski12,cggi18} are based on the \textit{Ring-Learning-With-Errors} (RLWE) problem~\cite{lwe, Lyubashevsky2010}. Such schemes operate over the polynomial ring $R_q {=}\mathbb{Z}_q[x]/(X^n + 1)$ (with $n$ the ring degree), and support \textit{Single-Instruction-Multiple-Data} (SIMD) operations, as they can pack vectors of data in a single plaintext/ciphertext. Moreover, such schemes require \textit{ad-hoc} cryptographic material to perform some operations (e.g., ciphertext-ciphertext multiplication or cyclic rotations of the vector of encoded data), in the form of \textit{evaluation keys}. An operation $\texttt{GenEvk}(C)$ generates the evaluation keys for a set of operations in $C$ using the secret key $sk$. Finally, we note that RLWE-based HE schemes can perform a finite number of operations on ciphertexts before decryption.

\subsubsection{Private Information Retrieval (PIR)}\label{subsub:backg_pir}
Let $\mathcal{DB}$ be a database with $N$ records $\{x_0,\dots,x_{N-1}\}$ replicated among $k {\geq} 1$ machines, controlled by an \textit{honest-but-curious} adversary $\mathcal{S}$~\cite{dolev1983security}. Let $\mathcal{U}$ be a user interested in privately retrieving the i-th element $x_i$ from one or more of the $k$ database replicas. Finally, let \pir{} be an information retrieval scheme consisting of three algorithms: (i) $\texttt{Query}(i)$ used by $\mathcal{U}$ to generate a query $\mathbf{Q}$ to retrieve $x_i$ from one or more database replicas; (ii) $\texttt{Answer}(\mathbf{Q}, \mathcal{DB})$ used by $\mathcal{S}$ to generate an answer $\mathbf{A}$ to $\mathbf{Q}$ using $\mathcal{DB}$; (iii) $\texttt{Extract}(\mathbf{A})$ used by $\mathcal{U}$ to extract $x_i$ from $\mathbf{A}$. We consider \pir{} a PIR scheme if, \pir{} is \textit{correct} (i.e., $\texttt{Extract}(\texttt{Answer}(\texttt{Query}(i), \mathcal{DB})){=}x_i,\; \forall x_i {\in} \mathcal{DB}$), and \textit{private}, i.e., $\mathcal{S}$ learns no information about $i$ from $\mathbf{Q} {=} \texttt{Query}(i)$. Recent PIR schemes~\cite{mulpir, Mughees2021OnionPIRRE, Menon2022SPIRALFH}) guarantee correctness and confidentiality, by encrypting queries with an HE scheme \he{} and by letting $\mathcal{S}$ compute the answers using homomorphic operations. $\mathcal{U}$ can then decrypt the result and retrieve the desired item.


\section{Related Work}\label{sec:related}

We review the related work on the security and privacy aspects of 5G (\ref{sub:5gsec}) and PIR schemes (\ref{sub:pir_related}). 

\subsection{5G Security and Privacy}\label{sub:5gsec}

There is sparse research focusing on security and privacy topics for lawful interception in 5G. To the best of our knowledge, this work is the first covering the identifier association with the LI\_HIQR interface. On the contrary, the security and privacy aspects of mobile networks such as 5G and its previous generations has been more broadly studied. Related to this work, the AKA protocol was modified in 5G to protect user privacy by including the SUCI (instead of the IMSI) during UE identification to the network. However, the AKA protocol has several flaws~\cite{LFM, AKA}. Similar vulnerabilities have been also identified for the recently introduced SUCI mechanism~\cite{nori}. Hong et al.~\cite{Hong2018} studied the privacy implications of the 5G-GUTI re-allocation, while Fraunholz et al.~\cite{Fraunholz2022_Probing} and Kune et al.~\cite{Kune2012} investigated the feasibility of IMSI probing as an attack against user privacy by relying on the infrequent change of 5G-GUTIs. Other attacks that allow UE identification and tracking are UE \textit{attach-request} fingerprinting~\cite{Fraunholz2022_fingerprinting, Fraunholz2023_fingerprinting, Shaik2019}, physical layer fingerprinting~\cite{Yin2021}, IMSI catching~\cite{Chlosta2021}, and signal overshadowing (AdaptOver)~\cite{AdaptOver}. Nie et al.~\cite{Nie2022} performed a measurement study about the state of implemented 5G security features in Beijing deployments, and they found that several crucial security enhancements of 5G are not in use. We refer the interested reader to Khan et al.~\cite{Khan2020} for an overview of user privacy challenges on the air interface and 5G countermeasures. Finally, ZipPhone~\cite{Sung2020} and Boeira et al.~\cite{Boeira2023} conducted research on how user privacy can be preserved if the CSP acts as an untrusted party.

\subsection{Private Information Retrieval}\label{sub:pir_related}

PIR schemes are categorized based on their security guarantees (Information-Theoretic PIR (IT-PIR) vs. Computational-PIR (cPIR)), or depending on the number of non-colluding databases (single-server PIR vs. multi-server PIR). Chor et al.~\cite{chor98} proved that the only single-server IT-PIR scheme is the \textit{naive} scheme that downloads the entire database: Such a scheme is highly impractical for real-world deployments, as it incurs a communication cost linear in the size of the database. Hence, many works focus on single-server cPIR schemes~\cite{mulpir, Mughees2021OnionPIRRE, Menon2022SPIRALFH}. Nonetheless, single-server cPIR schemes come with a high overhead, as the server performs computation linear in the size of the database~\cite{beiemel_pir_preprocess}. To overcome this limitation, single-server cPIR schemes \textit{with preprocessing}, pre-process the database in a (query-independent) offline phase to enable sublinear computational cost during the online phase~\cite{beiemel_pir_preprocess}. However, PIR schemes \textit{with-preprocessing} assume that the database is immutable; despite attempts to relax this assumption~\cite{incremental_preprocess_pir}, pre-processing the database is infeasible for real-world use-cases (e.g., for single-server PIR or PIR over key-value storage, also known as \textit{Keyword}-PIR). More recently, single-server cPIR schemes exploit the capabilities of modern homomorphic encryption schemes to increase their efficiency~\cite{pir_with_phi_hiding, pir_with_trapdoor_perm, mulpir, Mughees2021OnionPIRRE, Menon2022SPIRALFH}. In this work, we rely on single-server cPIR schemes, as in line with our threat model (i.e., a single ICF under the control of the adversarial CSP~\ref{sub:threat}), and we extend such protocols to make them practical and meet the operational requirements of LI.

\section{System and Threat Model}\label{sec:sys_threat_model}

We describe the system and threat models considered (\ref{sub:system} and~\ref{sub:threat}) as well as the requirements (\ref{sub:guarantees}) for practical and confidential LI on the 5G core.

\subsection{System Model}\label{sub:system}


\subsubsection{CSP}\label{subsub:csp_model}
We consider that the CSP and the LEA engage in a single query-response protocol, following the specifications for identity disclosure operations~\cite{3gpp.33.127, 3gpp.33.128}. We assume that there exists a single ICF in the 5G network, and a single IEF deployed in association with the AMF component. The IEF reports a new event every time a UE registration is successful or UE unregisters from the AMF~\cite{3gpp.33.127}.

\subsubsection{LEA}\label{subsub:lea_model}
We consider an operational scenario where interception happens \textit{on-field}, i.e., LEAs sniff traffic between UE and CSP with radio devices and query the CSP directly from the radio device using a mobile internet connection with possibly variable network conditions (e.g., depending on the area's congestion level or its location). The LEA network captures are incomplete versions of the events stored on the ICF. More concretely, we assume that an association/disassociation event is stored in the ICF as a tuple of data containing information such as the SUPI, concealed identifiers (e.g., SUCI, 5G-GUTI), start/end time of the associations between permanent and temporary identifiers, and location data (e.g., cell where the subscriber connected from). A LEA network capture is an \textit{incomplete} version of such tuple, e.g., missing the SUPI (which, in most cases, LEA aims to retrieve from the CSP), and potentially other information (e.g., temporary identifiers not observed \textit{over-the-air}, start/end validity time(s)). However, we note that at least one of the SUPI, SUCI, 5G-GUTI or 5G-TMSI is known by both LEA and CSP. We consider the following types of queries that LEAs can perform: by SUCI or by 5G-TMSI/5G-GUTI to associate temporary identifiers with a SUPI, or a \textit{reverse lookup} by SUPI in order to associate it with temporary identifiers used in the network (e.g., for real-time tracking). Finally, we note that as we focus on \textit{identity disclosure} (\ref{sub:li_5g}), the issuance of a warrant is generally not required (e.g., see the case of Switzerland~\cite{SR780.1, SR780.11}). As such, we consider that the only source of information leakage is the execution of the query-response protocol itself.

\subsection{Threat Model}\label{sub:threat}
We consider that the \ac{csp} is an untrusted entity following the \textit{honest-but-curious} model (equivalent to $\mathcal{S}$ in~\ref{subsub:backg_pir}). In particular, the CSP will try to gain information about \ac{lea} operations and the identities (and hence possible involvement in criminal proceedings) of its subscribers by observing the execution transcript of the LI query-response protocol over the LI\_HIQR interface. As such, all the components in the CSP domain (i.e., the IEF, ICF, IQF, etc.) are untrusted. We deem this threat model realistic for our use-case, as CSPs are required by law to cooperate with LEAs, hence, it would be detrimental for them to behave as \textit{active} adversaries (and deviate from the protocol). We consider the LEA as a trusted entity (similar to $\mathcal{U}$ in~\ref{subsub:backg_pir}), i.e., we do not consider the LEA acquiring information about the mobile subscribers as a privacy breach.

\subsection{Requirements}\label{sub:guarantees}
The LI architecture should ideally allow for identity resolution in a few seconds (i.e., to enable quick reaction from LEAs). The low latency requirement is particularly crucial for \textit{on-field} operations, like raids. Furthermore, the query-response protocol should provide the following guarantees:

\subsubsection*{1) Correctness}\label{subsub:correctness}
By running the protocol, the LEA should retrieve from the CSP all associations and disassociation events matching its queries.

\subsubsection*{2) Query Privacy}\label{subsub:query_privacy}
By running the protocol, the LEA queries should leak no information to the CSP about the retrieved events. As discussed in~\ref{sub:pir_related}, the computational overhead of classical cPIR schemes makes it impossible to balance the privacy of the query and the LI operational requirements. Hence, we further relax this guarantee as follows.

\subsubsection*{2') Query Quasi-Privacy}\label{subsub:query_quasi_privacy}
By running the protocol, the LEA queries should leak no \textit{non-voluntary} information to the CSP about the retrieved events. In other words, the LEA should be able to select how much information it wants to leak to the CSP about the retrieved events while obtaining guarantees that the CSP does not learn any information beyond that. We note that \textit{Query Quasi-Privacy} can correspond to \textit{Query Privacy}, i.e., the LEA can choose to run the protocol with no information leakage.

\section{A generic Information Retrieval scheme}
\label{sec:sparseWPIR}
We introduce \proto{}, our generic information retrieval scheme that fulfills the requirements of \textit{Correctness} and \textit{Query Quasi-Privacy} (\ref{sub:guarantees}). We first build our scheme starting from generic building blocks like RLWE-based HE and PIR schemes (\ref{sub:recursion_pir}--\ref{sub:selective_leakage}). Then, we provide a careful privacy analysis of the scheme (\ref{sub:leakage}).

\subsection{PIR with Recursion}
\label{sub:recursion_pir}
We build \proto{} on top of a generic PIR scheme \pir{} exploiting a RLWE-based HE scheme \he{} and a technique called \textit{Recursion}~\cite{Melchor2016XPIRP, sealpir, mulpir, Mughees2021OnionPIRRE, Menon2022SPIRALFH}. In the \textit{Recursion} model, the database $\mathcal{DB}$ is represented as a hyper-rectangle $\mathcal{D}$ in $d$ dimensions $\mathbf{K} {=} [K_0,\dots,K_{d-1}]$. Assuming that $\mathcal{DB}$ contains $N$ records, $\mathcal{D}$ is designed such that each of the $\prod_{i=0}^{d-1}K_i$ hyper-rectangle cells contains exactly $M$ records (where $M$ is the capacity of an RLWE plaintext). As such, a single cell of $\mathcal{D}$ is represented as a fully-packed RLWE plaintext (exploiting the SIMD capabilities discussed in \ref{subsub:backg_he}). We indicate a cell of $\mathcal{D}$ at coordinates $\{k_0, \dots, k_{d-1}\}$ as $\mathbf{x}_{k_0,\dots, k_{d-1}}$.


\subsection{Weakly-Private Information Retrieval}
\label{sub:vanilla_wpir}
We base \proto{} on a generalization of PIR from Information Theory called \textit{Weakly-Private Information Retrieval} (WPIR) \cite{wpircapacity}, which allows some fixed information leakage to reduce the overhead associated with the classical PIR protocol. In WPIR, the database $\mathcal{DB}$ is partitioned in $\eta$ partitions of size $N_\eta {=} \frac{N}{\eta}$. To retrieve the item $x_j$ at the index $j$ of the i-th partition, $\mathcal{U}$ builds the query $\mathbf{Q} {=} (\mathbf{\Tilde{Q}}, i) {\in} \mathcal{Q} {=} \mathcal{\Tilde{Q}} \times [0,...,\eta-1]$, where $\mathbf{\Tilde{Q}}$ is the realization of \pir{}.\texttt{Query}$(j)$. In more detail, $\mathcal{U}$ provides $\mathcal{S}$ a ``hint'' $\mathbf{h}$ about which partition the requested item belongs to and $\mathcal{S}$ runs the complete PIR protocol on a single partition of the database of size $N_\eta$. 

\subsection{Keyword PIR}
\label{sub:keyw_pir}
In real-world scenarios, databases are typically represented as key-value storage rather than indexed arrays, i.e., an item $x$ is associated with some keyword $w {\in} \mathcal{W}$ (where $\mathcal{W}$ is a large domain), rather than an index $i {\in} [0,\dots,N-1]$. Moreover, it may happen that many elements of $\mathcal{W}$ are not present in $\mathcal{DB}$ (i.e., $\mathcal{DB}$ is \textit{sparse} in $\mathcal{W}$). \textit{Keyword}-PIR tackles such database types~\cite{chor98_keywordpir}. However, existing \textit{Keyword}-PIR schemes, compared to their classical PIR counterparts, introduce significant communication and computation overhead~\cite{chor98_keywordpir, mulpir}, $\mathcal{O}(N)$ storage at the client side~\cite{chor98}, or assume static databases~\cite{sparse_pir}. We extend \proto{} to \textit{Keyword}-PIR without making specific assumptions nor specific optimizations (e.g., static databases, batch query optimizations, etc.), by adopting a similar solution to~\cite{mulpir}. In particular, we apply a mapping $\mathcal{H}{:}\mathcal{W} {\rightarrow} R^d$, with $\mathcal{H}$ a cryptographic hash function that associates a keyword $w$ related to a record, with a vector $\mathbf{s}$ of $d$ integers representing the coordinates of the hyper-rectangle cell that stores $w$. Hash collisions are not a concern, as we can pack up to $M$ records in a single cell represented as an RLWE plaintext. However, we need to correctly configure the hypercube dimensions ($K_0,\dots,K_{d-1}$), such that we can guarantee with high probability that a single cell of $\mathcal{D}$ can be always represented by a reasonably small number of plaintexts $\theta$. To achieve this, we exploit the results of Raab et al.~\cite{balls_n_bins} to estimate with overwhelming probability the worst-case number of collisions $M'$ in a single cell of $\mathcal{D}$ as a function of the number of cells $\prod_{i=0}^{d-1}K_i$, subject to the constraint $M' {\leq} \theta M$, $\theta {\geq} 1$. In particular, we model the problem of inserting $N$ events in $\prod_{i=0}^{d-1}K_i$ cells as a \textit{balls into bins problem}, and we follow~\cite{balls_n_bins} to find a value for each $K_i$ such that our constraint is met. We indicate an item associated to a keyword $w$ as $x_w$.

\subsection{WPIR on Sparse Databases with Selective Leakage}
\label{sub:selective_leakage}

\proto{} further expands the notion of WPIR by: (i) running WPIR on a sparse database, and (ii) allowing $\mathcal{U}$ to dynamically choose the set of partitions to run PIR on (parameterized by a leakage parameter $\epsilon$ which controls how much information it leaks to $\mathcal{S}$). We observe that we can dissect the dimensions of $\mathcal{D}$ and obtain a dynamic set of $\eta(\epsilon)$ partitions. Let's assume that $\mathcal{U}$ wants to retrieve the item $x_w$ associated with keyword $w$. $\mathcal{U}$ generates the selection vector $\mathbf{s} {=} \mathcal{H}(w) {=} [k_0,\dots,k_{d-1}]$ that maps $x_w$ to a cell $\mathbf{x}_{k_0,\dots, k_{d-1}}$ in $\mathcal{D}$ (\ref{sub:keyw_pir}). $\mathcal{U}$ generates a query $\mathbf{Q}$ as follows:
\begin{align}
    \mathbf{Q} = (\mathbf{\Tilde{Q}}, \mathbf{h}) \in \mathcal{Q} = \mathcal{\Tilde{Q}} \times (\mathbb{Z}_{K_0} \times \dots \times \mathbb{Z}_{K_{d_s(\epsilon)-1}})
\end{align}
with $d_s(\epsilon) {\in} [0,\dots,d]$. According to the leakage parameter $\epsilon$, $\mathcal{U}$ selectively leaks a hint to the server $\mathcal{S}$, represented by the vector $\mathbf{h} {=} [k_0, \dots, k_{d_s(\epsilon)}]$ containing the first $d_s(\epsilon)$ coordinates of $\mathbf{x}_{k_0,\dots, k_{d-1}}$ in $\mathcal{D}$. This is equivalent to running the WPIR construction of \ref{sub:vanilla_wpir} with $\eta {=} \prod_{i=0}^{d_s(\epsilon)-1} K_i$ and $N_\eta {=} \frac{N}{\prod_{i=0}^{d_s(\epsilon)-1} K_i}$. In other words, $\mathcal{S}$ runs the PIR protocol on the subset of $N_\eta$ elements of $\mathcal{D}$ with the first $d_s(\epsilon)$ coordinates matching what was leaked by $\mathcal{U}$ in $\mathbf{h}$. 

\subsection{Leakage Analysis}\label{sub:leakage}
We characterize the leakage of \proto{} by quantifying the uncertainty of the adversary trying to guess the item of interest $x_w$ when observing $\mathbf{Q}$, by using the \textit{min-entropy} metric from information theory:
\\
\begin{equation}
\begin{gathered}
    H_{\infty}(X {=} x_w | Q {=} \mathbf{Q}){=}\\
    -\log_2 \sum_{Q} Pr_{(Q)}(\mathbf{Q}) {\cdot} \max_{\substack{w \in \mathcal{W} | x_w \in \mathcal{DB}}} Pr_{(X|Q)}(x_w, \mathbf{Q}) \stackrel{(a)}{=}\\
    -\log_2 \sum_{Q} Pr_{(Q)}(\mathbf{Q}) \cdot \frac{1}{N_{\eta(\epsilon)}} \stackrel{(b)}{=}\\
    -\log_2(\frac{\|\mathcal{Q}\|}{\|\mathcal{Q}\|N_{\eta(\epsilon)}} )= \log_2 \frac{N}{\eta(\epsilon)}\stackrel{(c)}{=}
    H_{\infty}(X) - \rho
\end{gathered}
\label{eq:sparsewpir_entropy}
\end{equation}
where $\stackrel{(a)}{=}$ stems from the fact that the adversary can only guess uniformly at random in the set of elements $N_{\eta(\epsilon)}$, which corresponds to the partition of the database induced by running \proto{} with the hint produced with $\epsilon$; $\stackrel{(b)}{=}$ derives from the assumption that $\mathcal{U}$ looks for an item according to a uniform distribution, hence every query instance has the same probability of being generated over the alphabet $\mathcal{Q}$; $\stackrel{(c)}{=}$ follows from \textbf{Theorem 1} in \cite{wpircapacity}, with $\rho$ the \textit{Max Leakage} privacy metric which measures the bits of information leaked to $\mathcal{S}$ about the identity of the requested item $x_w$ by observing the transcript of a query execution $\mathbf{Q}$. We remark that, in \proto{}, \textit{min-entropy} has similar semantics to the privacy notion of \textit{k-anonymity} (a privacy metric used in real-world systems like ``Have I Been Pwned''~\cite{pwned_passwords}) with $k {=} \frac{N}{\eta(\epsilon)}$.
Overall, the leakage in \proto{} indicates how much information is leaked to $\mathcal{S}$ about the coordinates $\mathbf{s}$ of an item $x_w$ in $\mathcal{D}$. In the context of LI, this leakage can be practically interpreted by estimating the size of the anonymity set of all the possible keywords $w$ (e.g., network identifiers such as SUCI) associated with the query $\mathbf{Q}$ from the CSP's point of view.

\section{Practical and Confidential LI}
\label{sec:design}
We now describe \sys{}, a system which enables practical and confidential \ac{li} on the 5G core by leveraging on \proto{} (\ref{sec:sparseWPIR}) with LEA playing the role of $\mathcal{U}$ and the CSP the role of $\mathcal{S}$. We first present the workflow of \sys{} (\ref{sub:p3li5_agreement} and \ref{sub:ir_phase}) and then discuss additional techniques required to enable \proto{} in LI infrastructures (\ref{sub:dynamic_cache} and \ref{sub:multi_query}).

\subsection{Agreement Phase}\label{sub:p3li5_agreement}
Before starting with the information retrieval protocol, LEA and CSP execute the \textit{Agreement Phase} which consists of two algorithms:

\subsubsection{Context Generation}\label{subsub:ctx_setup}
The CSP setups a \textit{context} for the LI, by choosing the dimensions of the hyper-rectangle representation for the ICF. The details are shown in Algorithm~\ref{alg:ctx_setup}. We assume that the ICF already stores $N$ events. If the ICF is empty, the CSP can configure $N$ based on past statistics (e.g., the expected number of events).

\begin{algorithm}[H]
\caption{Context Generation}\label{alg:ctx_setup}
\scriptsize
\begin{algorithmic}[1]
\Require $N$ {:} Number of records; $B$ {:} record size; $d$ {:} Dimensions of $\mathcal{D}$ with $d \geq 2$; $n$ {:} Ring degree of \texttt{HE}; $\theta$ {:} A tolerance factor for the number of records that can be stored in a cell of $\mathcal{D}$.
\Ensure $\texttt{ctx} {=} (\mathbf{K}, d, n)$ (i.e., the \textit{context}).
\State Compute $M$ from $(B, n)$ \Comment{Max. \# of records stored in a cell of $\mathcal{D}$}
\State Estimate $\mathbf{K}$ such that $\prod_{i=0}^{d-1}K_i {\geq} \lceil\frac{N}{M}\rceil$ \Comment{e.g., $K_0 {=} \dots = K_{d-1}{=} \sqrt[d]{\lceil\frac{N}{M}\rceil}$}
\State Compute $M'$, i.e., worst case number of collisions in cell from $\mathbf{K}$ \Comment{as per (\ref{sub:keyw_pir})}
\If{$\theta {\cdot} M \geq M'$}
\State \Return $(\mathbf{K}, d, n)$
\EndIf
\State \Return $\bot$ 
\end{algorithmic}
\end{algorithm}
\subsubsection{Profile Generation}\label{subsub:client_setup}
LEA retrieves the \textit{context} from the CSP and creates a \textit{profile} for the LI execution. It selects the parameters for the HE scheme and it generates the evaluation keys required for the PIR protocol (see Algorithm~\ref{alg:client_setup} for details). 

\begin{algorithm}[H]
\caption{Profile Generation}\label{alg:client_setup}
\scriptsize
\begin{algorithmic}[1]
\Require $\texttt{ctx}$ {:} Context created by CSP; $\epsilon${:} Leakage parameter (\ref{sub:selective_leakage}); $\mathbb{P}$ {:} The space of all possible parameters for \texttt{HE}.
\Ensure $sk$ {:} Secret key for \he{}; $P_\epsilon {=} (p_\epsilon, \text{evk}_\epsilon)$ {:} Profile containing the parameters for \he{} and the evaluation keys required to execute \pir{}.

\State Extract a suitable set of parameters $p_\epsilon$ for \he{} from $\mathbb{P}$, given $\text{ctx}$ and $\epsilon$ \Comment{Select the ciphertext modulus $q$ for $R_q$ given $n$ from $\text{ctx}$}
\State $sk {\gets} \texttt{\text{\he{}}.KeyGen}(p_\epsilon)$
\State $\text{evk}_\epsilon {\gets}$ \he{}.\texttt{GenEvk}(\pir{}) \Comment{Generate eval. keys for the execution of \pir{}}
\State \Return $sk, P_\epsilon {=} (p_\epsilon, \text{evk}_\epsilon)$
\end{algorithmic}
\end{algorithm}

Note that LEA chooses the complete set of \texttt{HE} parameters, as these depend on $\epsilon$. In particular, higher values of $\epsilon$ allow for smaller values of $q$ (i.e., the modulus of the ring $R_q$ in \ref{subsub:backg_he}), which decrease the computational complexity of the HE operations. Moreover, we highlight that transferring \textit{profiles} incurs high communication, as the evaluation keys consist of several polynomials. However, the CSP can cache the \textit{profiles} after receiving them for the first time.

\subsection{Information Retrieval Phase}\label{sub:ir_phase}
After the LEA and CSP agree on the representation of the database as a hyper-rectangle $\mathcal{D}$, and LEA setups the desired cryptographic parameters and evaluation keys (\ref{sub:p3li5_agreement}), comes the \textit{Information Retrieval Phase}. We assume that both LEA and CSP employ a common encoding strategy $\mathcal{H}$ which maps events $\mathbf{e}$ and network captures $\mathbf{\Tilde{e}}$ to a fixed set of coordinates in $\mathcal{D}$. For example, the CSP can map an event $\mathbf{e}$ with an element $x_w$ (\ref{sub:keyw_pir}), by using an identifier (e.g., SUPI, or SUCI, 5G-GUTI, 5G-TMSI) as a keyword $w$, and by storing it in a cell $\mathbf{x}_{k_0,\dots,k_{d-1}}$. The LEA can do the same, as it is guaranteed that at least one identifier is known both by LEA and CSP (\ref{subsub:lea_model}). As per (\ref{sub:keyw_pir}), we use a hash-based mapping. \sys{} exposes two algorithms, $\texttt{Query}$ (Algorithm~\ref{alg:query}) executed by the LEA and $\texttt{Answer}$ (Algorithm~\ref{alg:answer}) executed by the CSP:

\begin{algorithm}[H]
\caption{Query}\label{alg:query}
\scriptsize
\begin{algorithmic}[1]
\Require $\texttt{ctx} {=} (\mathbf{K},d,n)$ {:} Context; $\epsilon$ {:} Leakage parameter; $P_\epsilon {=} (p_\epsilon, \text{evk}_\epsilon)$ {:} Profile; $sk$ {:} Secret key for \texttt{HE}; $\mathbf{\Tilde{e}}$ {:} Network capture.
\Ensure $\mathbf{Q}$ (Query).
\State Extract keyword $w$ from $\mathbf{\Tilde{e}}$
\State $\mathbf{k} {=} [k_0, \dots, k_{d-1}] {\gets} \mathcal{H}(w)$ \Comment{Map $w$ to coordinates in $\mathcal{D}$}
\State Determine $d_s(\epsilon) {\in} [0,\dots,d]$
\Comment{Select the amount of coordinates to leak based on $\epsilon$, e.g., $d_s(\epsilon) {=} \epsilon$, for $\epsilon {\in} [0,\dots,d]$.}
\State $\mathbf{h} {=} [k_0, \dots, k_{d_s(\epsilon)}]$
\State $\Tilde{\mathbf{k}} {\gets} [k_{d_s(\epsilon)},\dots,k_{d-1}]$
\State $\mathbf{\Tilde{Q}} {\gets} \text{\pir{}}.\texttt{Query}(\Tilde{\mathbf{k}})$
\State \Return $\mathbf{Q} {=} (\mathbf{\Tilde{Q}, \mathbf{h}})$
\end{algorithmic}
\end{algorithm}

\begin{algorithm}[H]
\caption{Answer}\label{alg:answer}
\scriptsize
\begin{algorithmic}[1]
\Require $\texttt{ctx}$ {:} Context; $\mathbf{Q} {=} (\mathbf{\Tilde{Q}, \mathbf{h}})$ {:} Query, $P_\epsilon$ {:} Profile sent by the LEA; $\mathcal{D}$ {:} Hyper-rectangle representation of $\mathcal{DB}$ (ICF).
\Ensure $\mathbf{A}$ (Answer).
\State $\Tilde{\mathcal{D}} {=} \{ \mathbf{x}_{k_0,\dots,k_{d-1}} \in \mathcal{D}\;|\;k_i {=} h_i,\; i \in [0,\dots, d_s(\epsilon)])$ \Comment{Partition the database using the hint $\mathbf{h} {=} [h_0,\dots,h_{d_s(\epsilon)}]$ (\ref{sub:selective_leakage}).}
\State \Return $\mathbf{A} {\gets} $ \pir{}.$\texttt{Answer}(\mathbf{\Tilde{Q}},\Tilde{\mathcal{D}})$ \Comment{Execute \pir{} on the partition of the database $\Tilde{\mathcal{D}}$}
\end{algorithmic}
\end{algorithm}

After receiving the answer $\mathbf{A}$ from the CSP, LEA uses \pir{}.\texttt{Extract}$(\mathbf{A})$ to obtain the cell $\mathbf{x}_{k_0,\dots,k_{d-1}}$ from $\mathcal{D}$, that contains the events relevant to its query.


\subsection{Handling Dynamic Cache}\label{sub:dynamic_cache}
A subtle problem that \sys{} needs to address, is the possibility of a \textit{context-mismatch} while executing the \texttt{Answer} algorithm (Alg.~\ref{alg:answer}). Indeed, as the ICF continuously inserts events from the IEF or deletes expired entries from the cache, the database representation can change after the LEA fetches the \textit{context} from the CSP (\ref{subsub:ctx_setup}). This forces the CSP to enlarge or shrink the dimensions of $\mathcal{D}$ which causes 2 problems: (i) LEA queries might be inconsistent with the current representation $\mathcal{D}$, requiring the CSP to send an updated \textit{context} such that LEA re-generates the query at the cost of an additional round-trip time; (ii) LEA is forced to regenerate and retransmit all the \textit{profiles}, including \textit{evk}, which is very costly (\ref{subsub:client_setup}). \sys{} tackles this problem by \textit{over-provisioning} the hyper-rectangle representation, i.e., by configuring the dimensions of $\mathcal{D}$ for a number of events larger than those currently stored in the ICF. This way, \sys{} accommodates database updates and shrinks the hyper-rectangle size only when the number of events goes under a certain threshold which can be tuned experimentally.

\subsection{Multi-Query Support}\label{sub:multi_query}

\sys{} can support queries by multiple keywords (e.g., by SUCI, (5G-)TMSI or SUPI, as per~\ref{sub:system}) by following three deployment options for the ICF:

\subsubsection{Single Hashing}
In this case, a single keyword (e.g., SUCI) is extracted from an event which is mapped accordingly to a single bin in the hyper-rectangle representation $\mathcal{D}$. Upon reception of a different query type (e.g., lookup by 5G-TMSI), a re-encoding process re-indexes all the database entries to a different bin with the correct keyword (e.g., 5G-TMSI). The storage complexity of this option is $\mathcal{O}(N)$ (where $N$ is the number of events in the ICF), and it incurs an additional linear cost of $\mathcal{O}(N)$ for re-encoding the entries on the fly before executing the \texttt{Answer} algorithm (\ref{sub:ir_phase}).

\subsubsection{Multiple Hashing} 
Different keywords are extracted from the same event (i.e., SUPI, SUCI and (5G-)TMSI) and the record is mapped to different bins according to the hash-mapping $\mathcal{H}$ and the keyword. This way, \sys{} supports successive queries of different types without additional work, in exchange for storage complexity: For $m$ different keywords (e.g., $m {=} 3$), \sys{} incurs a linear factor of $\mathcal{O}(mN)$.

\subsubsection{Distributed Single Hashing}
Following this option, \sys{} employs the single hashing technique in a distributed fashion, i.e., it sets up $m$ ICF components (with $m$ being the number of supported look-up types) and the IEF sends events to all ICFs. Each ICF employs the single-hashing technique on a different keyword. This allows \sys{} to support queries for different keywords simultaneously, at the cost of maintaining more machines.

\subsection{Extension to Personal Identifiable Information}\label{sub:personal_info}
While the scope of \sys{} is mainly the retrieval of association information between permanent and short-term identifiers used at the network-level, we note that it can be extended to support the retrieval of subscribers' personal identifiable information (PII). For instance, PII could be included in the representation of events at the ICF (e.g., at the cost of performing a join operation using the SUPI with the internal CSP database storing subscribers' PII and assigned SUPI, and of a small storage overhead to store the extra information). Alternatively, \sys{} can be used a second time to query the CSP database storing subscribers' PII, using the retrieved SUPI as a common keyword between the LEA and the CSP.

\section{Experimental Evaluation}
\label{sec:evaluation}
We present our implementation details (\ref{sub:implementation}) and the evaluation benchmarks of \texttt{SparseWPIR} (\ref{sub:sparsewpir_eval}). Then, we describe our proof-of-concept implementation of \sys{} (\ref{sub:eval_p3li5}). Our code is open-source~\cite{p3li5, sparsewpir, pyli5, open5gsLI, ueransimLI}.

\begin{figure}[t]
\centering
\includegraphics[width=\columnwidth]{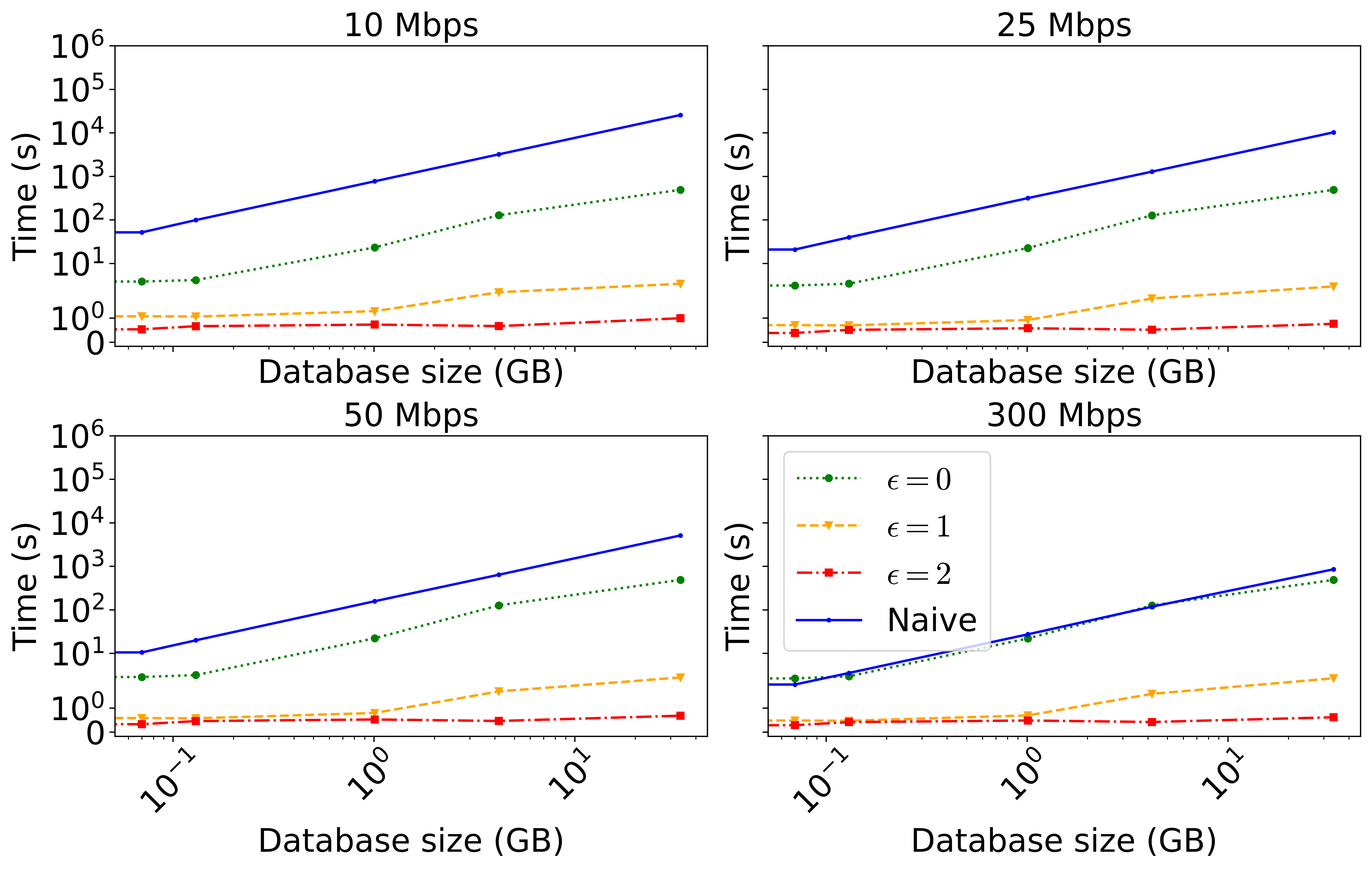}
\caption{\proto{} end-to-end latency under different network conditions and database sizes (in log scale), using different leakage parameters ($\epsilon$).}
\label{fig:sparsewpir_latency}
\end{figure}

\subsection{Implementation Details}\label{sub:implementation}

\subsubsection{Hardware}
All experiments are performed on a machine running Ubuntu 20.04, with 256 GB DDR4 RAM and an 2.80GHz Intel(R) Xeon(R) Gold 6242 processor with 64 cores.

\subsubsection{Software}
We implement \proto{} in Go, using the \texttt{Lattigo} homomorphic encryption library~\cite{lattigo}. To simulate variable network conditions, we rely on the Go \texttt{Latency} package. For the underlying PIR scheme, we reimplemented the \textit{MulPIR} scheme~\cite{mulpir} in Go, with a multithreaded execution of the \texttt{Answer} algorithm (Alg.~\ref{alg:answer}). \textit{MulPIR} uses the BFV HE scheme~\cite{FV12,bfv_rns,Brakerski12}, which we instantiate with parameters that achieve 128-bit security~\cite{he_standard}. Nonetheless, we remark that it is easy to integrate more recent PIR schemes~\cite{Mughees2021OnionPIRRE,Menon2022SPIRALFH} into \proto{}. For PIR \textit{Recursion}, we use $d{=}3$ dimensions, and we generate \proto{} hints such that $d_s(\epsilon) {=} \epsilon$. Moreover, we represent the database as a hyper-rectangle where $K_0 {=} K_1 {=} \dots {=} K_{d-1}$ (i.e, a hyper-cube).

\subsection{\proto{} Evaluation}\label{sub:sparsewpir_eval}
We evaluate the algorithmic part of \sys{} at scale and in a typical IR setting by benchmarking the online phase of \proto{}. In particular, we measure the end-to-end latency of the \texttt{Query} (Alg.~\ref{alg:query}), \texttt{Answer} (Alg.~\ref{alg:answer}) and \texttt{Extract} algorithms. We exclude the offline part (context generation (Alg.~\ref{alg:ctx_setup}) and profile generation (Alg.~\ref{alg:client_setup}) as its latency cost is negligible (in light of~\ref{sub:dynamic_cache}), however, we stress that transferring \textit{profiles} can be costly in terms of communication as the size of a \textit{profile} ranges between $20-40$MB). We test \sys{} on databases populated with synthetic data generated according to the expected byte size of events stored in the ICF~\cite{3gpp.33.127}, and of various sizes up to $\sim$34GB, which we deem to be a realistic estimation of the worst-case size of the ICF. In particular, following the results of Leo et al.~\cite{network_mobility}, we consider that IEF events are generated following a Poisson process with an intensity of $\lambda_{\text{Poisson}}{=}0.0006\ \text{registrations/s}$ at peak-hour, a maximum retention time for the ICF of $t_{\text{max}}{=}54$ minutes~\cite{3gpp.33.127}, and a short-term caching time of $t_{\text{short}}{=}\frac{t_{\text{max}}}{2}$ minutes. Then, the database size is estimated as $N_{\text{sub}} {\cdot} \lambda_{\text{Poisson}} {\cdot} t_{\text{short}}(s) {\cdot} b$, with $N_{\text{sub}}$ the number of network subscribers and $b$ the byte length of an event in the ICF. For a large CSP like Verizon ($\sim$143M users as of 2023~\cite{verizon_subs}), assuming  $b{\approx}250$B~\cite{3gpp.33.127}, an estimated worst-case size for the ICF is $\sim$34.7GB. We also use different values for the leakage parameter $\epsilon$, and we simulate variable bandwidth conditions (according to 5G reference values~\cite{3gpp.22.261}) relevant to different interception scenarios: a congested network (10Mbps), broadband access in a crowded urban area (25Mbps), access in a remote rural area or from a moving vehicle (50Mbps), and access in a densely populated urban area (300Mbps).

\begin{figure}[t]
\centering
\includegraphics[width=\columnwidth]{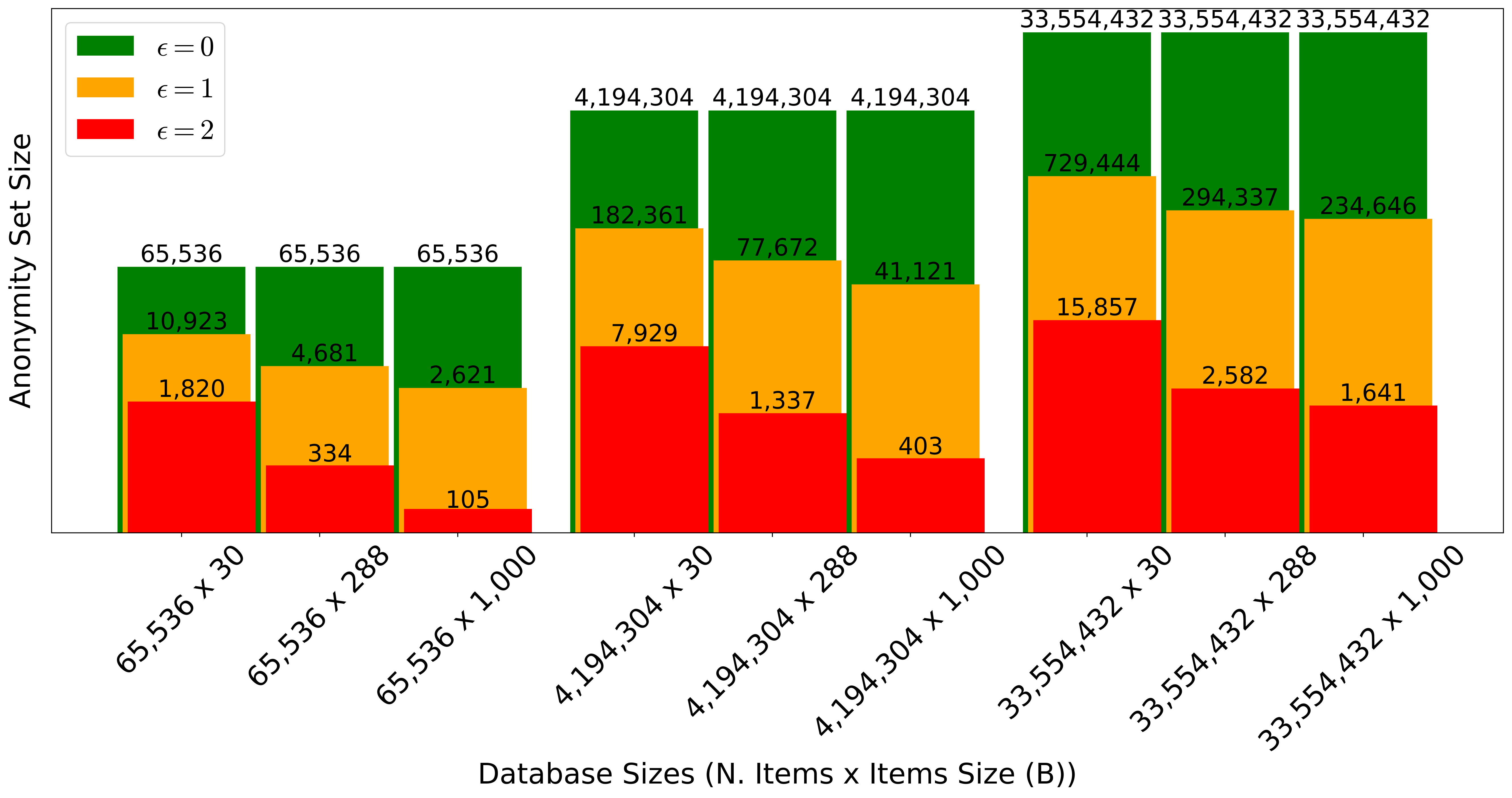}
\caption{Size of the anonymity sets induced by \proto{} for different database configurations and leakage parameters ($\epsilon$). $\epsilon{=}0$ (green) offers full privacy (i.e., $k{=}N$ in the context of $k$-anonymity~\ref{sub:leakage}).}
\label{fig:sparsewpir_leak}
\end{figure}

Figure~\ref{fig:sparsewpir_latency} shows the \proto{} latency benchmarks for increasing database sizes, with the \textit{naive} PIR as a baseline. First, we note how trading the communication cost of downloading the whole database (\textit{naive} PIR) for the computational cost of \proto{} becomes more beneficial in operational scenarios with limited bandwidth. Nonetheless, (the fully private version of) \proto{} (green) performs almost always better than the \textit{naive} scheme (blue), even for small databases with high bandwidth availability. Then, we highlight the scalability of \proto{}, whose latency grows linearly with the database size. Finally, we observe that the fully confidential version achieves a speedup of up to two orders of magnitude over the \textit{naive} version, depending on the available bandwidth. Then, the quasi-private versions with moderate and high information leakage allow for a speedup ranging from $10^2\times$ to $10^4\times$ (moderate) and from $10^3\times$ to $10^5\times$ (high). 

\begin{figure}[t]
\centering\begin{minipage}{.35\columnwidth}
\centering
\includegraphics[width=\columnwidth]{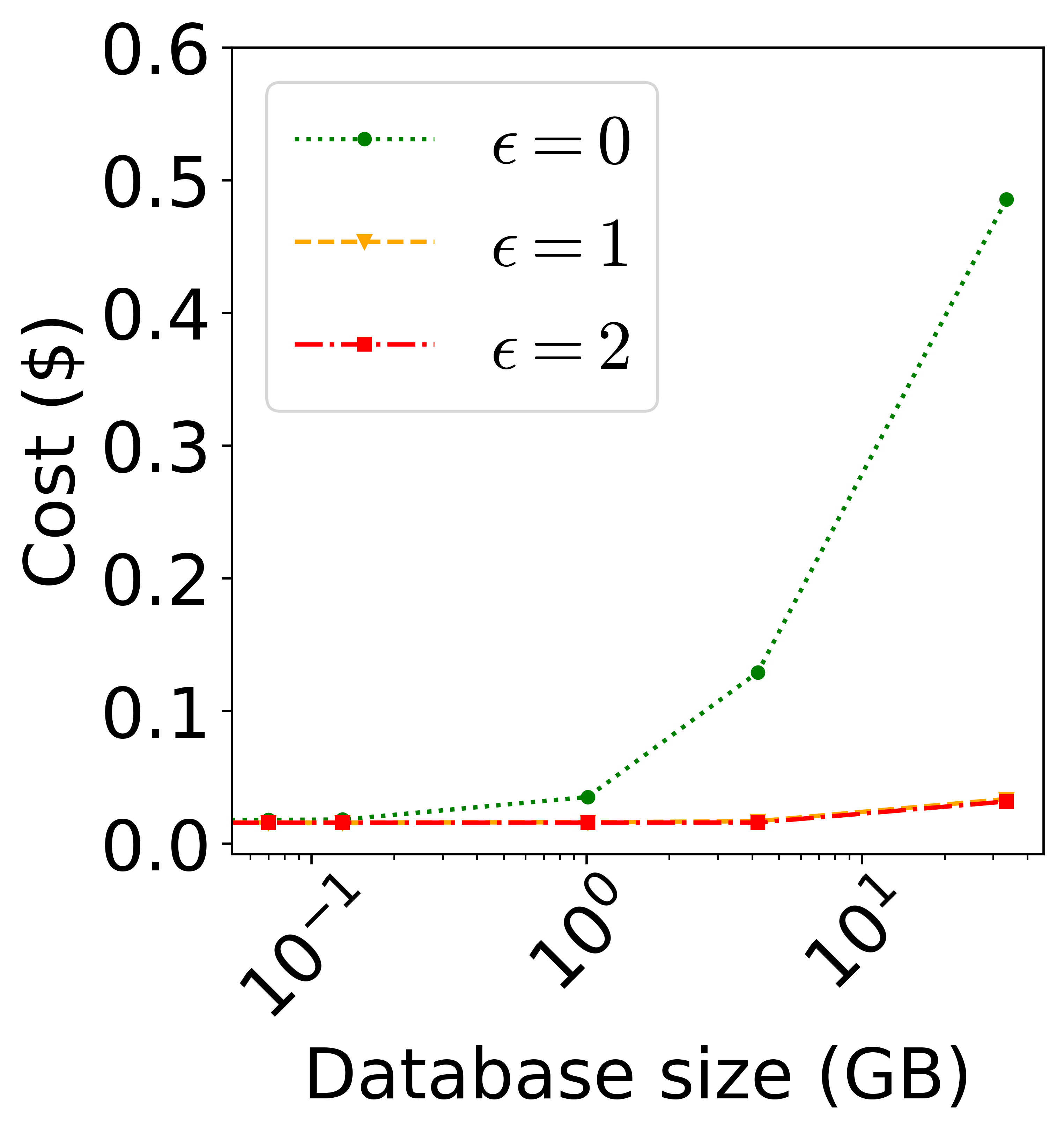}
\caption{Costs of \proto{} \texttt{Answer} protocol on an AWS \textit{m6g.16xlarge} machine (\textit{eu-central-2}) \cite{aws}.}
\label{fig:sparsewpir_cost}
\end{minipage}
\hfill
\begin{minipage}{.63\columnwidth}
\centering
\includegraphics[width=\columnwidth]{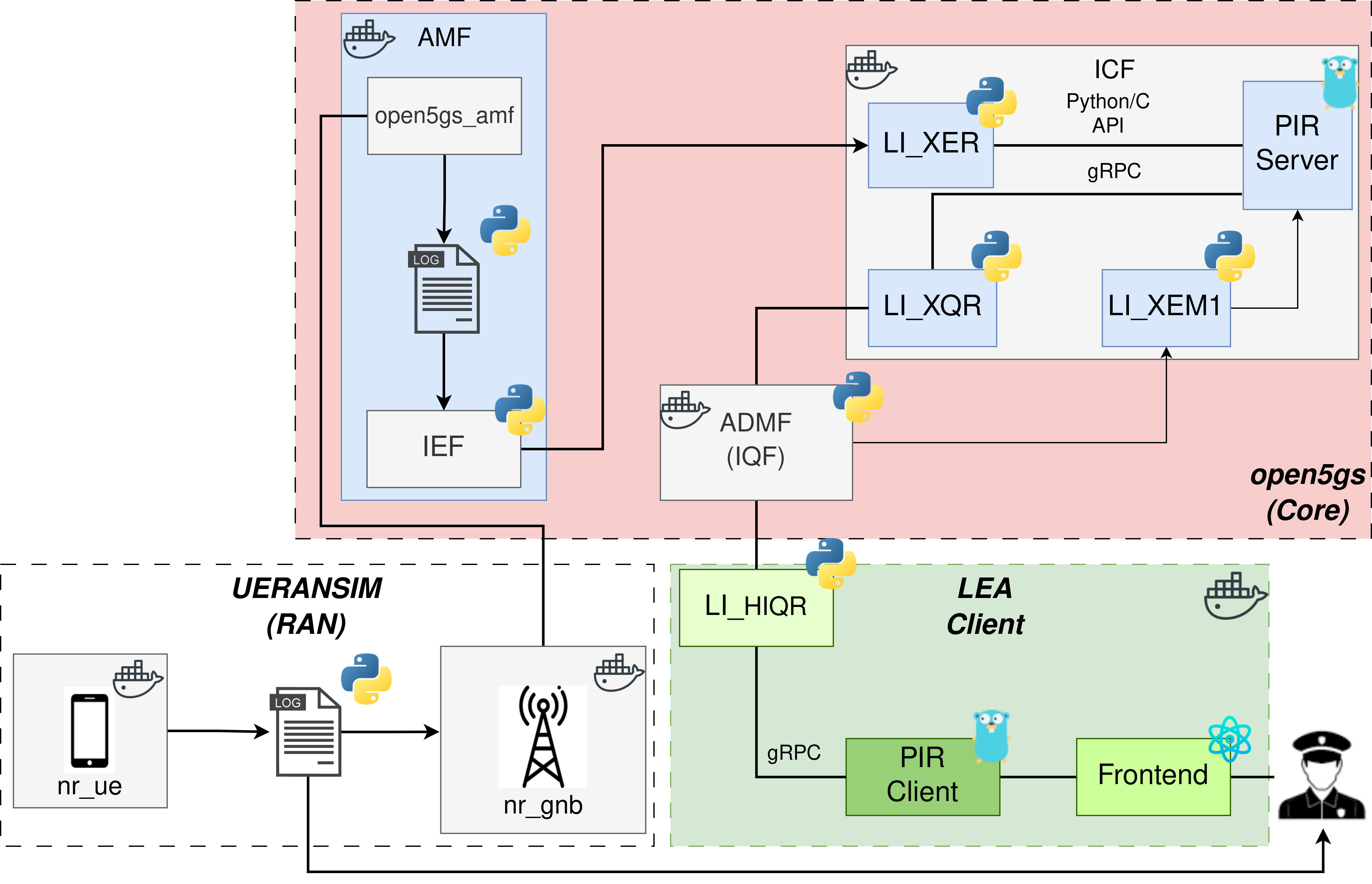}
\caption{\sys{} proof-of-concept implementation in a virtualized environment.}
\label{fig:p3li5_architecture}
\end{minipage}
\end{figure}

Figure~\ref{fig:sparsewpir_leak} provides insights about the information leakage of \proto{}, by showing how the leakage parameter $\epsilon$ affects the anonymity set size of the item retrieved by a query for databases with different numbers of entries, and entries of variable byte sizes (\ref{sub:leakage}). As expected, the higher the leakage parameter, the smaller the anonymity set size. Moreover, we observe that the size of the anonymity set is not only impacted by $\epsilon$, but also by the byte size of the items in the database: the smaller their byte size, the more items that can be packed in a cell of the hyper-rectangle, thus, enlarging the anonymity set. Finally, we evaluate the deployment costs of our \proto{}-enabled ICF on the cloud (e.g., AWS \cite{aws}). We take into account the pricing for both network and CPU time. Figure~\ref{fig:sparsewpir_cost} shows that a fully confidential query is always executed with less than 50¢ (USD cents), while quasi-private ones run with less than 5¢.

\subsection{\sys{} Proof-of-concept}\label{sub:eval_p3li5}

We build a proof-of-concept for \sys{} on a simulated interception task that resolves the identities of 100 subscribers by SUCI. We build a LI infrastructure for 5G on top of custom forks~\cite{open5gsLI, ueransimLI} of well-known open-source projects to simulate the 5G Radio Access Network (using UERANSIM~\cite{ueransim}) and the 5G Core network (using open5gs~\cite{open5gs}). We also implement a minimalistic LI infrastructure from scratch in Python~\cite{pyli5}. For ease of deployment, each component (UEs, base stations, 5GC NFs, LI components, LEA) is deployed on a separate Docker container. We dockerize open5gs and UERANSIM, using an open-source project~\cite{dockerized} which we modify to include their \proto{}-enabled versions~\cite{p3li5, sparsewpir}. Finally, we develop and integrate a new set of messages into the specifications~\cite{3gpp.33.126, 3gpp.33.127, 3gpp.33.128, 3gpp.03.221} to support \sys{}: for the LI\_HIQR interface, we define messages carrying only \proto{} encrypted queries/answers, suppressing all spatio-temporal metadata, and we follow a similar approach for LI\_XQR. A demonstration of \sys{} proof-of-concept (showing the \sys{} dashboard during the interception task) can be found \href{https://youtu.be/WpHpDL9VDm8}{online}.

\section{Conclusion}\label{sec:conclusion}

In this work, we presented \sys{}, a novel system that enables Law Enforcement Agencies (LEAs) to perform Lawful Interception (LI) on the 5G core network~\cite{3gpp.21.915} while protecting the confidentiality of their operations against untrusted Communication Service Providers (CSPs). \sys{} leverages on a novel information retrieval scheme, \proto{}, based on \ac{pir} and its weakly private version \ac{wpir}, hence achieving a tradeoff between privacy and performance. Our experimental results show that by selectively disclosing some bounded information, LEAs can query the CSP identifiers cache (ICF) and resolve identities with a performance speedup of up to $100\times$ over the fully private version of the protocol with a classical PIR scheme. Thus, \sys{} can adapt to several operational scenarios and scale to large ICFs ($\sim$34GB). Overall, \sys{} is the first solution to address the privacy issues raised by the requirement for LI on the 5G core.



\section{Acknowledgments}
We would like to express our gratitude to Prof. James Larus for supervising the master thesis from which this work originated, and the Very Large Scale Computing Lab (VLSC) at EPFL for funding its publication at IEEE CNS 2023. We would also like to thank Jean-Pascal Chavanne (Swiss Federal Dept. of Justice and Police, Switzerland) for providing useful insights about the 5G LI ecosystem and the legal framework in Switzerland, and Sylvain Chatel (EPFL) for reviewing a draft of this paper.
\bibliographystyle{IEEEtran}
\bibliography{IEEEabrv,bib}

\end{document}